\documentclass{article}
\usepackage{spconf,amsmath,amssymb,amsfonts,graphicx,hyperref}

\usepackage{booktabs}

\ninept

\newcommand{\lidffn}{LID\_FF}

\newcommand{\wntaffn}{WNTA64\_FF}

\title{Robust Accent Identification via Voice Conversion and Non-Timbral Embeddings}
\name{Rayane Bakari$^{1,2}$, Olivier Le Blouch$^{1}$, Nicolas Gengembre$^{1}$, Nicholas Evans$^{2}$}
\address{
    $^{1}$ Orange Innovation, France \\
    $^{2}$ EURECOM, Sophia Antipolis, France \\
    \{mohamedrayane.bakari, olivier.leblouch, nicolas.gengembre\}@orange.com, evans@eurecom.fr
}

\begin{document}
%
\maketitle
\begin{abstract}


Automatic accent identification (AID) remains a challenging task due to the complex variability of accents, the entanglement of accent cues with speaker traits, and the scarcity of reliable accent-labelled data. To address these challenges, we propose a speaker augmentation strategy using voice conversion (VC), with which we generate additional training data by converting original training utterances into different speaker voices while preserving accentual cues. For this purpose, we select two recent VC systems and evaluate their capability to preserve accent. 
Alternatively, we also explore the use of non-timbral embeddings in AID, for their ability to convey accent information among other non timbral cues.
The effectiveness of both methods is demonstrated on the GenAID benchmark, achieving a new state-of-the-art F1-score of 0.66, compared to the previous score of 0.55. Beyond AID, we show that non-timbral embeddings enable accent-controlled Text-to-Speech, producing high-fidelity speech with accurate accent transfer.

\end{abstract}

\section{Introduction}
\label{sec:intro}
Accent is a key aspect of spoken language, reflecting geographic, social, and cultural variation, encompassing phonemic, phonetic, rhythmic, and structural features~\cite{wells1982accents}. Automatic accent identification (AID) has attracted attention for application in speech recognition, spoken language understanding, and sociolinguistic studies. Despite advances in related areas like language identification (LID) and speaker recognition, accent recognition remains a challenging task.

A major challenge is the complex variability of accents, often linked to a speaker's native language. Unlike speaker differences, accent variation relies on subtle phonetic and phonological cues. Studies~\cite{bafna25_interspeech} show that LID models misclassify non-native (L2) speech as the speaker’s native or related language, illustrating the difficulty of separating accent from language and other speaker traits. The lack of large-scale datasets and the entanglement of speaker and accent cues further hinder the learning of robust models that generalise to unseen speakers.

Recent deep learning approaches~\cite{shi2021accentedenglishspeechrecognition} improve AID accuracy but may struggle to disentangle speaker and accent cues, highlighting the need for speaker-invariant representations and greater data diversity.
Data augmentation is a key strategy to address data scarcity in many speech tasks~\cite{fischbach25_interspeech}. Conventional approaches~\cite{kim21c_interspeech,park19e_interspeech} improve robustness but may distort accent cues. In this work, we use voice conversion (VC) as a speaker augmentation technique to tackle two key challenges simultaneously. First, by generating additional training data from multiple target speakers, VC directly alleviates data scarcity. Second, speaker augmentation helps disentangle speaker-related and accent information, since the same accent can be expressed in a number of different voices. This property makes VC particularly well-suited for AID tasks, as it allows models to focus on accentual patterns without being confounded by speaker-specific characteristics. Importantly, effective augmentation requires a VC system that preserves accentual traits while modifying speaker identity.
Among recent VC systems, Retrieval-based Voice Conversion (RVC)\footnote{\url{https://github.com/RVC-Project/Retrieval-based-Voice-Conversion-WebUI}} and k-Nearest Neighbors VC (kNN-VC)~\cite{baas23_interspeech} have shown strong capabilities in disentangling timbre from linguistic content, but their use for data augmentation in AID remains unexplored. 

To the best of our knowledge, this is the first work to analyse the behaviour of specific VC systems for designing speaker augmentation strategies to AID. Specifically, we evaluate RVC and kNN-VC for their ability to alter timbre while preserving accent. We also explore the use of specialized embeddings (LID~\cite{lyth2024naturallanguageguidancehighfidelity}, non-timbral~\cite{gengembre2024disentangling}) to enhance AID by improving speaker invariance.
Experiments on the GenAID benchmark~\cite{accentbox} show significant improvements in generalization to unseen speakers. 

The main contributions of this work are as follows~:
\begin{itemize}

    \item we present the first systematic analysis of VC systems (RVC, kNN-VC)  for speaker diversity and accent-preserving augmentation~;
    \item we design task-aware augmentation strategies that leverage VC properties~;
    \item we use specialized embeddings to enhance accent recognition by promoting speaker invariance~;
    \item we improve state-of-the-art results on GenAID benchmark with unseen speakers by 10 points~;
    \item we demonstrate the novel use of non-timbral embeddings in accent-controlled Text-to-Speech (TTS), showing their effectiveness for accent transfer.
\end{itemize}

\section{Related Works}
\label{sec:rel_work}
\subsection{Accent Identification}
AID shares similarities with  LID \cite{langage_identi, pratap2023scalingspeechtechnology1000} and speaker identification~\cite{TIRUMALA2017250, REYNOLDS199591}, but is challenged by data scarcity, limited accent diversity, and imbalanced speaker–accent distributions. Early AID systems used context-dependent HMMs~\cite{texeira} or formant frequency-based GMMs~\cite{1544415}. The AESRC2020 benchmark~\cite {shi2021accentedenglishspeechrecognition} enabled systematic comparisons but is no longer publicly available. Recent AID approaches leverage large-scale self-supervised speech models to extract accent-discriminative embeddings. For instance, \cite{lyth2024naturallanguageguidancehighfidelity} proposed an AID pipeline that uses embeddings extracted from a LID model~\cite{pratap2023scalingspeechtechnology1000}, classifies them with a simple linear layer, achieving strong performance; however, evaluation on unseen speakers was unclear, leaving open the risk of speaker–accent entanglement.

Large-scale accent datasets, such as CommonAccent~\cite{zuluagagomez23_interspeech} and GLOBE~\cite{wang24b_interspeech} derived from Common Voice~\cite{cvoice}, lack reliable speaker-level accent metadata, complicating generalization. In GLOBE, some metadata were auto-generated using HuBERT~\cite{hubert_ref} with ECAPA-TDNN~\cite{desplanques2020ecapa}, a model that may introduce label noise.  To address evaluation bias from overlapping speakers,~\cite{accentbox} introduced speaker-disjoint train/test splits based on CommonAccent and released GenAID\footnote{\url{https://github.com/jzmzhong/GenAID/tree/GenAID}}, an accent classifier that achieved a 0.55 F1-score on unseen test speakers, which is claimed to be the state-of-the-art performance.

\subsection{Speaker-Invariant Representations and Disentanglement}
A key challenge in AID is speaker–accent entanglement, where models inadvertently capture speaker-specific cues instead of accent features. Speaker-invariant representation learning addresses this by extracting embeddings that retain specific information while minimizing speaker identity cues. Adversarial training~\cite{sun2018domain, accentbox} is commonly used, jointly optimizing an encoder with a speaker classifier to suppress timbre information, improving robustness to unseen speakers~\cite{shi2021accentedenglishspeechrecognition}. Alternative disentanglement approaches explicitly separate timbral and non-timbral cues. For example, \cite{gengembre2024disentangling} introduced complementary embeddings for timbral and non-timbral features while~\cite{melechovsky2024accentconversiontexttospeechusing} proposed a multi-level VAE with an accent classifier to disentangle accent and speaker embeddings. \cite{jain18_interspeech} applied a TDNN with a bottleneck layer to produce frame-level accent embeddings, facilitating more effective downstream classification.

\subsection{ Voice conversion for Data Augmentation: }
Beyond representation learning, data augmentation is another strategy to improve AID. Traditional augmentations such as SpecAugment~\cite{park19e_interspeech}, SpecMix~\cite{kim21c_interspeech}, time-stretching, pitch shifting, additive noise, and synthetic speech generation~\cite{9053008}, increase model robustness by introducing variability at the spectral or temporal level, but can distort speaker and accent cues. Voice conversion (VC) offers controlled modification of speaker identity while preserving target attributes~\cite{vc_adv}, making it particularly promising for AID.
VC-based augmentation has been applied to multiple contexts: children’s speech recognition~\cite{shahnawazuddin20_interspeech}, speaker-independent keyword recognition~\cite{vc_keywspot}, speaker recognition under adverse conditions~\cite{tao2024voiceconversionaugmentationspeaker}, etc. 
RVC and kNN-VC~\cite{baas23_interspeech} achieve high-quality conversions with minimal distortion of the underlying linguistic attributes.  Notably,~\cite{fischbach25_interspeech} recently showed RVC benefits for low-resource dialect classification, though their work was limited to a single VC method.


\section{Timbre-Accent Disentanglement}
\label{sec:analysis}

\begin{figure*}[ht]
    \centering
    \includegraphics[width=1.\textwidth]{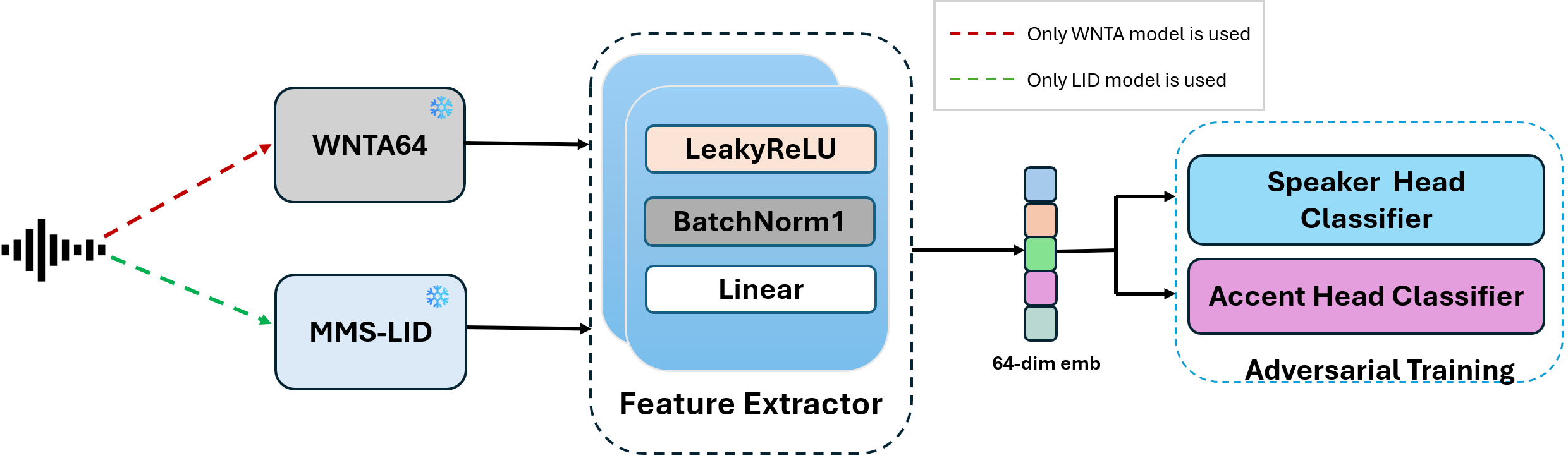}
    \caption{Architecture of the AID model. The model takes as input either LID embeddings extracted from a pretrained MMS-LID model or non-timbral WNTA64 embeddings. The embeddings are fed into a feed-forward classifier with two heads: an accent classification head and a speaker classification head, the latter being used during adversarial training with KL regularization.}
    \label{fig:wnta_architecture}
\end{figure*}

While RVC is used as a framework for disentangling different aspects of speech~\cite{gengembre2024disentangling, fischbach25_interspeech}, we hypothesize that more recent VC models, such as kNN-VC, may be better suited to this task. Thus, we have used objective metrics to evaluate whether VC systems can modify speaker timbre while preserving non-timbral attributes, and specifically the accent.
We therefore compare RVC and KNN-VC along two main axes:

\begin{enumerate}

   \item \textbf{Timbre modification}: We compute cosine similarity between speaker embeddings extracted using ECAPA-TDNN\footnote{\url{https://huggingface.co/speechbrain/spkrec-ecapa-voxceleb}}. These embeddings are well-established for capturing timbral characteristics~\cite{gengembre2024disentangling}, making them a suitable choice for evaluating speaker similarity. We calculate the speaker similarity between the source and converted speech, and between the target and converted speech.
   \item \textbf{Accent preservation}: We estimate retention of accent using the GenAID model~\cite{accentbox}. Since classification accuracy is limited by the accent predictor’s imperfections, we also use the Accent Embedding Cosine Similarity (AECS)~\cite{inoue2025macstmultiaccentspeechsynthesis}, which provides a continuous, more nuanced measure of similarity between the source and converted speech.
\end{enumerate}


For these evaluations, we perform voice conversion on the \emph{unseen test subset} of the GenAID benchmark, which comprises 13 accents, with 100 utterances per accent from distinct speakers. For each source utterance, four random target speakers are selected from dataset for conversion.

Table~\ref{tab:speaker_similarity_avg} presents the average speaker similarity scores and accent preservation results for RVC and kNN-VC. Both systems achieve low similarity with the source speaker, indicating a shift away from the original speaker identity. However, kNN-VC demonstrates a significantly higher similarity to the target speaker, suggesting more effective timbre conversion compared to RVC. It also shows a modest improvement in accent retention, as measured by both classification accuracy and AECS. For reference, the mean AECS between random utterance pairs is 0.80, serving as a baseline threshold; values above this indicate that accent cues are preserved.
Overall, these results indicate that both RVC and kNN-VC are effective at modifying timbre while preserving accent, with kNN-VC performing slightly better, consistent with our hypothesis. 

\begin{table}[!ht]
    \caption{Timbre modification and accent preservation for RVC and kNN-VC on unseen GenAID test set. Lower speaker similarity with the source and higher with the target indicates better timbre conversion. Higher accent accuracy and AECS indicate better retention of the original accent.}
    \vspace{3mm}  
    \label{tab:speaker_similarity_avg}
    \centering
    \begin{tabular}{l|cc|cc}
    \hline
     \textbf{VC System} &  \multicolumn{2}{c|}{\textbf{Speaker Similarity}} & \multicolumn{2}{c}{\textbf{Accent prediction}}\\
    \cmidrule(lr){2-3} \cmidrule(lr){4-5}
     & Source↓ & Target↑ & Acc↑ & AECS↑\\
    \hline
    RVC     & 0.19 {\scriptsize$\pm$ 0.10} & 0.48 {\scriptsize$\pm$ 0.12} &  48.19 & 0.84\\
    kNN-VC  & \textbf{0.18} {\scriptsize$\pm$ 0.10} & \textbf{0.70} {\scriptsize$\pm$ 0.08} & \textbf{50.12} &  \textbf{0.90} \\
    \hline
    \end{tabular}
\end{table}

\section{Proposed Approaches}
\label{sec:method}

Following the GenAID training setup, we aim to improve the AID model through better input representations and training data diversity. Two complementary strategies were explored: (1) speaker and accent-disentangled data augmentation using VC, and (2) accent-specific input representations via specialized embeddings.

\subsection{Voice conversion-Based Data Augmentation}
Building upon our previous analysis showing that specific VC systems such as RVC and kNN-VC can effectively mask timbre while preserving accentual cues, we employ one of these methods at a time to augment the training data.  For each original training utterance, we generate two additional voice-converted versions using the selected VC system, with target speaker identities randomly drawn from the LibriTTS train-clean-100 subset~\cite{zen19_interspeech}. This process increases speaker variability in the training set 
while minimally affecting the underlying accent, thereby encouraging the model to learn speaker-invariant accent features. 

\subsection{Specialized Embedding-Based Representations}


Beyond data augmentation, we investigate the use of specialized embeddings to improve accent classification by leveraging latent accent-relevant cues. Inspired by~\cite{lyth2024naturallanguageguidancehighfidelity}, we first extract embeddings from a pretrained LID model\footnote{\url{https://huggingface.co/facebook/mms-lid-256}}~\cite{pratap2023scalingspeechtechnology1000}, which are known to encode language-specific phonetic patterns potentially useful for accent discrimination. In parallel, we employ WNTA64 embeddings~\cite{gengembre2024disentangling} extracted from a pretrained model\footnote{\url{https://huggingface.co/Orange/Speaker-wavLM-pro}} designed to disentangle timbral and non-timbral cues. These embeddings are taken from an intermediate layer that emphasizes prosody and accentual information while minimizing speaker-specific timbre. Compared to other embeddings like x-vector or HuBERT embeddings, WNTA64 embeddings may offer a more explicit separation between accentual patterns and speaker timbre, making them especially suitable for our accent identification task~\cite{gengembre2024disentangling}.


The architecture used in our AID models is described in \autoref{fig:wnta_architecture}. LID or WNTA64 embeddings are extracted from the input signals, and then passed through three linear layers with batch normalization and ReLU activation, with output sizes 256, 128 and 64. As in~\cite{accentbox}, the resulting 64-dim embedding is fed into two heads dedicated to accent and speaker classification respectively, in an adversarial process. The two model variants are denoted as \lidffn{} and \wntaffn{}, with reference to their Feed Forward nature.

For adversarial training, the speaker classifier is trained to correctly predict the speaker, while the feature extractor and the accent head are optimized together to predict the accent and to fool the speaker classifier, encouraging the embeddings to be speaker-invariant through the Kullback-Leibler divergence (denoted as KL) as a penalty. The overall training loss is defined as:
\begin{equation}
\label{eq:loss}
\mathcal{L}_{\text{total}} = \mathcal{L}_{\text{accent}} + \lambda \cdot \text{KL} \left( \mathbf{p}_{\text{speaker}} \;\|\; \mathcal{U} \right)
\end{equation}
where $\mathcal{L}_{\text{accent}}$ is the cross-entropy loss for accent classification, $\mathbf{p}{_\text{speaker}}$ is the predicted speaker distribution, $\mathcal{U}$ is a uniform distribution over speaker classes, and $\lambda$ is a coefficient controlling the strength of the speaker-invariance constraint.

When optimized, the KL loss tends to bring the speaker prediction distribution close to a uniform distribution, thus reducing the amount of speaker-related information brought to the speaker head, i.e. contained in the 64-dim embedding. 

\section{Experimental Setup}
\label{sec:exp_setup}
All experiments follow the GenAID benchmark protocol~\cite{accentbox}, which is based on the CommonAccent corpus (derived from Common Voice v17.0). The data is split into training, validation, and test sets, with strict separation of speakers to ensure that evaluation reflects generalization to unseen speakers rather than memorization of speaker–accent mappings.

All proposed AID systems are trained for 10 epochs using distinct learning rates: 1e-4 for the accent classifier and 1e-5 for the auxiliary speaker classifier. This setup stabilizes adversarial training by allowing accent classification to dominate optimization, while still enforcing a gradual removal of speaker-specific information from the representations. 

The value of $\lambda$ in \autoref{eq:loss}, which also controls the adversarial process, is set to $0.1$.


\section{Results and Discussion}
\label{sec:result_disc}
Table~\ref{tab:aid_results} outlines the performance of AID systems on the unseen speaker subset of the GenAID benchmark. We compare the baseline GenAID model, its VC-augmented variants (using either RVC or kNN-VC), and the feed-forward classifiers \lidffn{} and \wntaffn{}, with and without VC-based augmentation. 

Retraining the baseline GenAID classifier with RVC-  or kNN-VC-augmented data consistently improves performance on unseen speakers. For example, RVC augmentation increases accuracy from 0.56 to 0.61, while kNN-VC yields an even higher accuracy of 0.66. These results confirm that VC-based augmentation effectively enhances generalization by increasing speaker diversity in the training set while preserving the original accent, thereby encouraging the model to learn speaker-invariant representations. Moreover, the superior performance of kNN-VC highlights that the effectiveness of augmentation is closely tied to the quality of timbre-accent disentanglement: better VC behavior leads to greater improvements in AID performance.

It is worth noting that a similar and complementary experiment involving both sets of augmented data, from RVC and from kNN-VC, did not result in any further performance enhancement. This indicates that increasing the volume of training data is not necessary, neither is the combination of different voice conversion methods. 
Furthermore, the adversarial speaker head is incorporated to ensure fair comparison and also promote timbre disentanglement. An incremental ablation study in~\cite{accentbox} demonstrates that this component is essential for learning speaker-invariant, accent-consistent representations, thereby justifying its inclusion in the proposed system.

Among the systems \#1, \#4 and \#7, the latter (\wntaffn{}) achieves the best overall performance, achieving \textbf{a 10-point improvement} (absolute, in percent) on accuracy over the baseline \#1, and nearly the same improvement when considering the f1-score. When combined with any a posteriori VC augmentation, it maintains the same high performance. This shows that isolating non-timbral or accent-relevant features, whether using WNTA64 or LID embeddings, is highly effective for accent generalization. VC-based augmentation adds only marginal gains for \lidffn{} or \wntaffn{}, as these embeddings already encode accent-specific information while minimizing speaker-specific timbre. As a result, the additional speaker variability introduced by VC-based augmentation has a negligible impact, since the model is already largely invariant to speaker identity. In other words, when using specialized embeddings, the benefits of VC augmentation are naturally reduced compared to models relying on raw features.

\autoref{tab:aid_results} also shows that the LID-based models do not perform as well as the WNTA64-based ones, suggesting that some of the accent information is not captured in these embeddings.

Overall, these results indicate that both VC-based data augmentation and specialized, non-timbral embeddings lead to similar improvements on unseen speakers, validating the effectiveness of our two approaches. 

\begin{table}[ht]
   \small 
    \setlength{\tabcolsep}{4pt}    
    \caption{Accent identification results. Tick (\checkmark) indicates which VC-based augmentation is applied.}
    \vspace{3mm}  
    \label{tab:aid_results}
    \begin{tabular}{l|cc|cccc}
    \hline
       \textbf{AID Systems} & \multicolumn{2}{c|}{\textbf{VC Aug}} 
     & \multicolumn{4}{c}{\textbf{Unseen Spks↑}} \\
      \cmidrule(lr){2-3} \cmidrule(lr){4-7}  & {\footnotesize RVC} & {\footnotesize kNN-VC} & prec & rec & f1  & acc \\

     \hline 
    \#1 GenAID$_{baseline}$ & - & - & 0.63 & 0.56 & 0.55 & 0.56 \\
    \#2 GenAID$_{RVC}$ & \checkmark & - & \textbf{0.72} & 0.61 & 0.60 & 0.61 \\
    \#3 GenAID$_{knnVC}$& - & \checkmark &  0.70 & \textbf{0.66} & 0.65 & \textbf{0.66} \\
    \hline
    \#4 \lidffn{}  & - & - &  0.57 & 0.58 & 0.57 & 0.58 \\
    \#5 \lidffn{}$_{RVC}$ & \checkmark & - &  0.57 & 0.58 & 0.58 & 0.58 \\
    \#6 \lidffn{}$_{knnVC}$ & - & \checkmark & 0.59 & 0.58 & 0.57 & 0.58 \\
    \hline
    \#7 \wntaffn{} & - & - &  \textbf{0.66} & \textbf 0.66 & \textbf{0.66} & \textbf{0.66} \\
    \#8 \wntaffn{}$_{RVC}$ & \checkmark & - &  0.65 & 0.65  & 0.65 & 0.65  \\
    \#9 \wntaffn{}$_{knnVC}$& - & \checkmark & 0.66 & \textbf{0.66} & \textbf{0.66} & \textbf{0.66}\\
    \bottomrule
    \end{tabular}
\end{table}

\section{Accent-Controlled TTS}
\label{sec:tts}
In addition to our main experiments, we evaluate the practical utility of non-timbral embeddings compared to dedicated accent representations in an accent-controlled TTS.
Building on the AccentBox framework~\cite{accentbox}, we implemented two TTS systems: the original AccentBox using
GenAID accent embeddings, and a modified version where we replaced the GenAID embeddings with WNTA64 embeddings, which may isolate accent-specific features without speaker information. 
For timbre control, both are also conditioned on WTA embeddings\footnote{\url{https://huggingface.co/Orange/Speaker-wavLM-tbr}} from \cite{gengembre2024disentangling} to avoid potential confounding effects related to speaker identity.

We evaluated the quality of accent propagation in generated speech using the GenAID model, measuring how accurately the system produces the intended accent. While this approach provides a consistent evaluation, it is important to note that using GenAID to assess the generated speech may introduce a bias in favor of GenAID embeddings, since the classifier is inherently optimized for the features and representations produced by the GenAID model itself. 
The evaluation dataset contains 720 generated utterances, built from 5 phonetically rich sentences and conditioned on 12 VCTK speakers, 2 per each of 6 accents.
It includes 144 unique voice–accent combinations (12 timbres × 12 accents).

Results in Table~\ref{tab:tts_results}, show that using WNTA64 embeddings improves accent control across all accents. This indicates that WNTA64 embeddings effectively encode accent information and can be used to generate speech with high accent fidelity. Overall, these findings demonstrate that these embeddings are not only useful for accent recognition but also valuable for controlling accent in speech synthesis, broadening their application in speech technology. 
\begin{table}[ht]
\small
\centering
\caption{Accent classification performance on TTS outputs conditioned on GenAID vs WNTA embeddings.}
\vspace{3mm}  
\label{tab:tts_results}
\begin{tabular}{l|ccc|ccc}
\hline
& \multicolumn{3}{c|}{\textbf{WTA + GenAID}} & \multicolumn{3}{c}{\textbf{WTA + WNTA}} \\
\cmidrule(lr){2-4} \cmidrule(lr){5-7}
Accent & Prec & Rec & F1 & Prec & Rec & F1 \\
\hline
US & 0.56 & 0.48 & 0.52 & 0.62 & 0.51 & \textbf{0.56} \\
Australian & 0.42 & 0.50 & 0.45 & 0.50 & 0.46 & \textbf{0.48} \\
South Asian & 1.00 & 0.06 & 0.11 & 1.00 & 0.11 & \textbf{0.20} \\
English & 0.41 & 0.81 & 0.55 & 0.47 & 0.92 & \textbf{0.62} \\
Scottish & 0.77 & 0.53 & 0.63 & 0.97 & 0.50 & \textbf{0.66} \\
Irish & 0.67 & 0.33 & 0.44 & 0.74 & 0.64 & \textbf{0.69} \\
\hline
\end{tabular}
\end{table}


\section{Conclusions}
\label{sec:concl}
In this work, we demonstrate the effectiveness of using voice conversion and specialized embeddings to improve robust accent identification for unseen speakers. By systematically analysing VC systems like RVC and kNN-VC, we showed that targeted data augmentation or the use of non-timbral, speaker-invariant embeddings significantly enhances accent recognition performance, achieving state-of-the-art performance with a 0.66 F1 score on 13-accent classification for unseen speakers. Additionally, our downstream experiments with accent-controlled TTS confirmed that these embeddings can effectively encode accent information, enabling high-fidelity accent synthesis. Future work will focus on developing TTS systems aimed at creating more inclusive and adaptable speech technologies, capable of accurately reproducing a wide range of accents for diverse applications.

\vfill\pagebreak



\let\oldthebibliography\thebibliography
\let\endoldthebibliography\endthebibliography
\renewenvironment{thebibliography}[1]{
  \begin{oldthebibliography}{#1}
    \setlength{\itemsep}{1pt}  
    \setlength{\parskip}{0pt}  
    \setlength{\parsep}{0pt}   
     \ninept                   
  }{
  \end{oldthebibliography}
}

\bibliographystyle{IEEEbib}
\bibliography{refs}

\end{document}